\documentclass[%
reprint,
superscriptaddress,
amsmath,amssymb,aps,
]{revtex4-2}
\usepackage{amsmath}
\usepackage{graphicx}
\usepackage{bm,physics}
\usepackage{color}
\usepackage{overpic}
\usepackage{tcolorbox}
\usepackage{hyperref}
\hypersetup{pdfborder={0 0 0}}

\usepackage{physics,empheq,bm}
\usepackage{xcolor}
\usepackage{textcomp}
\usepackage{mathcomp}

\begin{document}
\preprint{APS/123-QED}

\title{Attraction-controlled torque organization and rotational states in frictional granular matter}

\author{Kiwamu Yoshii}
\email{qyoshii@rs.tus.ac.jp}
\affiliation{Department of Applied Physics, Tokyo University of Science, 6-3-1 Nijuku, Katsushika-ku, Tokyo, 125-8585, Japan}

\date{\today}
\begin{abstract}
In this study, we numerically investigate how interparticle attraction affects stress transmission, torque organization, and particle rotation in a two-dimensional frictional granular material.
Although the attraction is a central force and exerts no direct torque on the particles, it strongly modifies the contact network and particle rotation.
Attraction produces a low-rate stress state and selects the high-stress state near the shear-thickening regime of the dry system, while leaving the high-rate thickened state almost unchanged.
By decomposing the shear stress, we find that the direct attractive contribution accounts for only a fraction of the total stress, while the repulsive contribution is strongly enhanced.
Attraction increases the coordination number and torque amplitude and can reverse the local torque-sign correlation.
At low packing fractions and intermediate shear rates, a stress-collapse state emerges with strongly suppressed torque and rotation despite a finite contact network. These results show that attraction controls macroscopic rheology through reorganization of the frictional contact network.
\end{abstract}
\maketitle

\section{Introduction}
\label{sec:intro}

The shear rheology of dense disordered systems, such as colloids,
suspensions, and granular materials, is closely related to the jamming
transition~\cite{liu1998jamming,ohern2003jamming}
and is a central topic in soft-matter physics.
For frictional particles, tangential contact forces couple translational
and rotational motion, so particle torque and spin provide microscopic
information about the frictional contact network that transmits stress.
In dry frictional granular materials, changes in torque organization have
been reported across different rheological states.
For example, Ref.~\cite{rahbari2021fluctuations}
showed that like-sign torque correlations on the low-stress branch
disappear across discontinuous shear thickening (DST).
More generally, frictional contact-network formation plays an important role
in shear thickening and jamming in both dense suspensions
and dry granular materials~\cite{seto2013discontinuous,
wyart2014discontinuous,mari2014shear,otsuki2011critical,
bi2011jamming,grob2014jamming,
grob2016rheological,otsuki2020shear,
d2025topological}.

The rheology of attractive granular materials has been studied over a wide range of packing fractions in wet granular systems
\cite{
herminghaus2013wet,mitarai2012granular,
roy2017general,vo2020additive}
and cohesive granular systems
\cite{rognon2008dense,gu2014rheology,
shi2020steady,mandal2020insights,mandal2021rheology,takada2018rheology,yoshii2023rheology,irani2014impact,irani2016athermal,
zheng2016shear,koeze2018sticky,Koeze2020Elasticity,
yoshii2025mechanical}.
Attraction generates cohesive stress at low shear rates
\cite{rognon2008dense,gu2014rheology,mandal2020insights}
and can induce yielding, shear banding, and hysteresis
\cite{mandal2021rheology,singh2019yielding}.
Nonmonotonic flow curves and discontinuous shear thinning were reported for frictionless adhesive dispersions in Ref.~\cite{irani2019discontinuous}, while yielding and shear jamming in cohesive frictional suspensions were investigated in Ref.~\cite{singh2019yielding}.
However, compared with the macroscopic rheology, much less is known about how attraction modifies torque organization and particle rotation within frictional contact networks.
It is also important to distinguish the stress directly carried by attraction from changes in stress transmission caused by contact-network reorganization.

The short-range attraction introduced in this study is a central force acting along the normal direction and therefore exerts no direct torque on the particles.
Thus, any change in particle torque or rotation induced by attraction must arise indirectly through reorganization of the frictional contact network.
We therefore compare the total shear stress with the repulsive and attractive virial contributions to distinguish the direct contribution of attraction from the change in stress transmission through the contact network.
We further characterize the microscopic rotational states using the coordination number, torque amplitude, torque-sign correlation between contacting particles, and particle spin.
Our purpose is to clarify how short-range attraction reorganizes the frictional contact network and thereby changes stress distribution, torque organization, particle rotation, and macroscopic flow states.

For this purpose, we numerically investigate a two-dimensional frictional granular system by varying the attraction strength, packing fraction, and shear rate.
In Sec.~\ref{sec:model}, we describe the model and numerical method.
In Sec.~\ref{subsec:adhesive}, we present the adhesive stress plateau, high-stress branch selection, and stress collapse observed in the flow curves.
In Sec.~\ref{subsec:rateresp}, we examine the responses of the contact network, particle torque, and rotation.
In Sec.~\ref{subsec:budget}, we compare the total shear stress with its repulsive and attractive contributions.
In Sec.~\ref{subsec:states}, we compare the real-space structures and single-particle statistics of representative states.
In Sec.~\ref{subsec:map}, we summarize the simulation results in state maps.
Finally, we discuss and summarize our results in Sec.~\ref{sec:discussion}.

\begin{figure}[htb]
  \centering
  \includegraphics[width=0.9\linewidth]{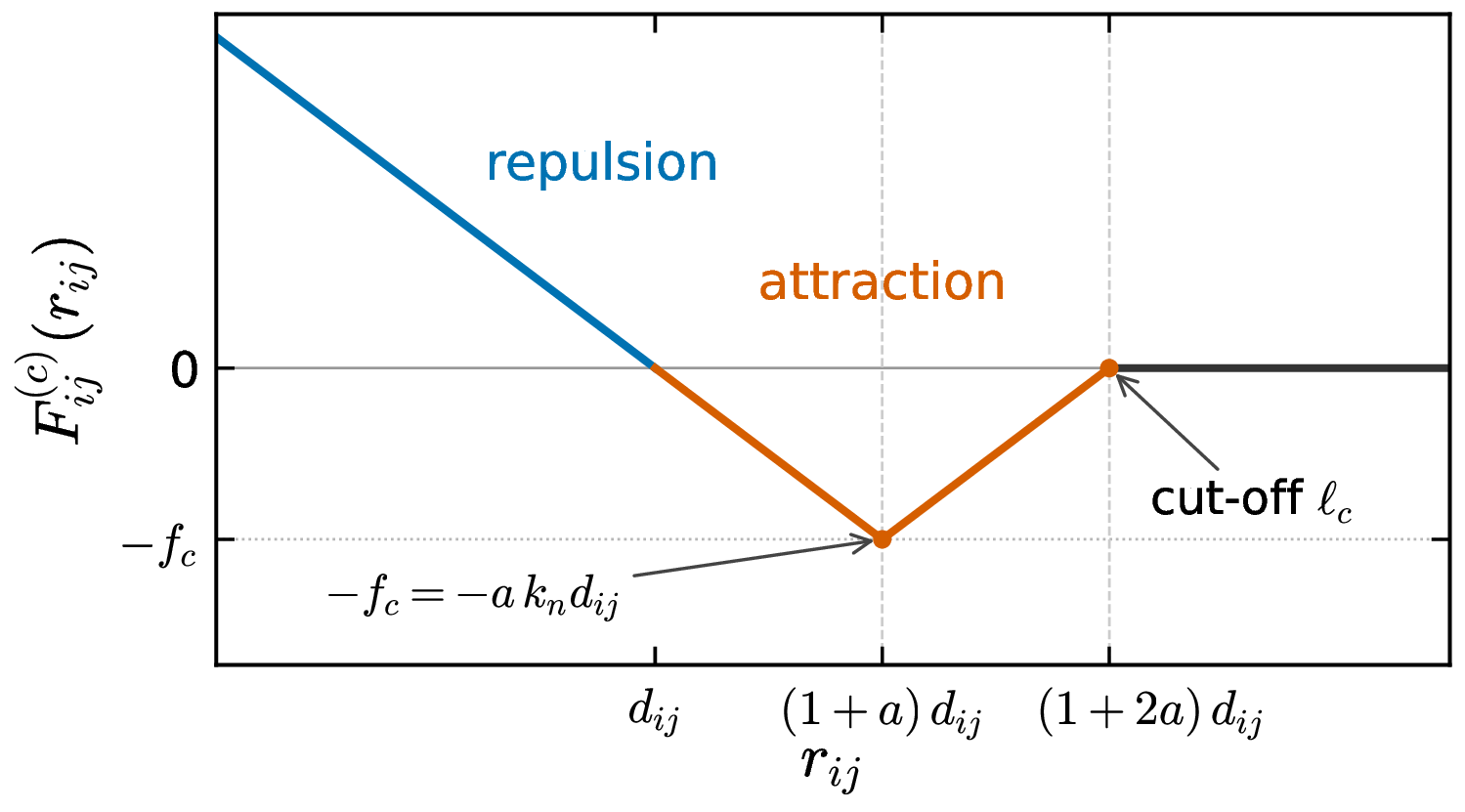}
  \caption{
Schematic of the normal force $F^{\rm (c)}_{ij}(r_{ij})$ [Eq.~\eqref{eq:fc}].
The attractive branch continuously extends the elastic spring from $r_{ij}=d_{ij}$, reaches
$-f_{c,ij}=-a k_{\rm n}d_{ij}$ at $r_{ij}=(1+a)d_{ij}$,
and vanishes at the cutoff $\ell_{c,ij}=(1+2a)d_{ij}$.
}
  \label{fig:model}
\end{figure}

\section{Model and numerical method}
\label{sec:model}
We consider a two-dimensional system of frictional disks.
To avoid crystallization, we use a 50:50 binary mixture of particles with diameters $d_0$ and $1.4d_0$.
The mass of a small particle is denoted by $m_0$, and the particle mass is proportional to its area, $m_i\propto d_i^2$.
Thus, $m_{\rm S}=m_0$ and $m_{\rm L}=1.4^2m_0$.
The translational motion of particle $i$ is governed by
\begin{align}
   m_i \dfrac{\dd^2 \bm{r}_i}{\dd t^2}
   = \sum_{j\neq i}
   \left[
   F_{ij}^{\rm (n)}\bm{n}_{ij}
   +F_{ij}^{\rm (t)}\bm{t}_{ij}
   \right]
   \label{eq:eom}
\end{align}
where $\bm r_{ij}=\bm r_i-\bm r_j$,
$r_{ij}=|\bm r_{ij}|$,
and $\bm n_{ij}=\bm r_{ij}/r_{ij}$.
Here, $\bm t_{ij}$ is the tangential unit vector perpendicular to $\bm n_{ij}$ \cite{luding2008cohesive}.
$F_{ij}^{\rm (n)}$ and $F_{ij}^{\rm (t)}$ denote the scalar components of the normal and tangential forces, respectively.
The angular velocity $\omega_i$ obeys
\begin{align}
   I_i \dfrac{\dd \omega_i}{\dd t}
   =-\dfrac{d_i}{2}
   \sum_{j\neq i}
   \left(\bm n_{ij}\times F_{ij}^{\rm (t)}\bm t_{ij}\right)_z
   \equiv \tau_i
   \label{eq:eom_rot}
\end{align}
where $I_i=m_i d_i^2/8$ is the moment of inertia of a disk and $\tau_i$ is the frictional torque acting on particle $i$.
Because the normal force does not appear in Eq.~\eqref{eq:eom_rot}, the interparticle attraction introduced below exerts no direct torque on the particles.

The normal force is given by the sum of conservative and dissipative forces,
$F_{ij}^{\rm (n)}=F_{ij}^{\rm (c)}+F_{ij}^{\rm (d)}$.
With $d_{ij}=(d_i+d_j)/2$, we introduce
$u_{ij}=a d_{ij}$,
$r_{m,ij}=d_{ij}+u_{ij}$,
$\ell_{c,ij}=d_{ij}+2u_{ij}$,
and
$f_{c,ij}=k_\mathrm{n}u_{ij}$,
where $a\ge0$ is a dimensionless attraction parameter.
The conservative force is given by
\begin{empheq}[left={F^{\rm (c)}_{ij}(r_{ij})=\empheqlbrace}]{alignat=2}\label{eq:fc}
    & k_\mathrm{n}(d_{ij}-r_{ij})
    &\quad (r_{ij}<r_{m,ij}),\\
    &-k_\mathrm{n}(\ell_{c,ij}-r_{ij})
    &\quad (r_{m,ij}\le r_{ij}<\ell_{c,ij}),\\
    &0
    &\quad (r_{ij}\ge\ell_{c,ij}).
\end{empheq}
The first branch gives a linear repulsive force for $r_{ij}<d_{ij}$ and is extended with the same slope into the attractive region for $d_{ij}<r_{ij}<r_{m,ij}$.
Hereafter, we define particles with $r_{ij}<d_{ij}$ as being in contact; thus, a contact corresponds to an overlapping pair with elastic repulsion.
The force reaches its minimum value
$-f_{c,ij}=-a k_\mathrm{n}d_{ij}$
at $r_{m,ij}=(1+a)d_{ij}$ and then returns to zero at
$\ell_{c,ij}=(1+2a)d_{ij}$ [Fig.~\ref{fig:model}].
In the figure and below, we use the maximum attractive force for a pair of small particles,
$f_c\equiv a k_\mathrm{n}d_0$,
as the label for each run.
The dry system corresponds to $a=0$, and we examine
$a=10^{-5},10^{-4},10^{-3}$.
Note that, in the present model, changing $a$ changes both the maximum attractive force and the interaction range.

With the normal relative velocity
$v_{ij}^{\rm (n)}=(\dot{\bm r}_i-\dot{\bm r}_j)\cdot\bm n_{ij}$,
the normal dissipative force is given by
\begin{align}
   F_{ij}^{\rm (d)}
   =-\eta_\mathrm{n}v_{ij}^{\rm (n)}\Theta(d_{ij}-r_{ij})
   \label{eq:fd}
\end{align}
and acts only on overlapping particle pairs.
Here, $\Theta(x)$ is the Heaviside function satisfying
$\Theta(x)=1$ for $x\ge0$ and $\Theta(x)=0$ otherwise.

The tangential force also acts only on overlapping particle pairs.
For the Coulomb threshold, we use the normal contact force consisting of the elastic repulsion and normal dissipation, excluding the attractive force.
The tangential force\cite{luding2008cohesive} is given by
\begin{align}
   F_{ij}^{\rm (t)}
   &=-\mathrm{sgn}(\delta_{ij})\Theta(d_{ij}-r_{ij})
   \nonumber\\
   &\quad\times\min\!\left[
   k_\mathrm{t}|\delta_{ij}|,
   \mu_p
   \left|
   k_\mathrm{n}(d_{ij}-r_{ij})+F_{ij}^{\rm (d)}
   \right|
   \right]
   \label{eq:ft}
\end{align}
,
where $k_\mathrm{t}$ is the tangential spring constant,
$\mu_p$ is the interparticle friction coefficient,
$\mathrm{sgn}(x)$ denotes the sign of $x$, and
$\min(x,y)$ denotes the smaller of $x$ and $y$.
The tangential displacement is given by
$\delta_{ij}=\int v_{ij}^{\rm (t)}\,\dd t$
during contact, with
$v_{ij}^{\rm (t)}=(\dot{\bm r}_i-\dot{\bm r}_j)\cdot\bm t_{ij}
-(d_i\omega_i+d_j\omega_j)/2$.
Here, $\delta_{ij}$ is reset when $r_{ij}\ge d_{ij}$.
Thus, only the conservative attractive force acts in the attractive shell
$d_{ij}\le r_{ij}<\ell_{c,ij}$,
while the tangential and normal dissipative forces vanish.
The interparticle attraction therefore neither appears directly in Eq.~\eqref{eq:eom_rot} nor enters the Coulomb threshold in Eq.~\eqref{eq:ft}.

We place $N=8192$ particles in a square box of side length $L$.
Periodic boundary conditions are imposed in the $x$ direction, and Lees--Edwards sliding periodic boundary conditions \cite{evans2008non} are imposed in the $y$ direction.
A constant shear rate $\dot\gamma$ is applied with the flow direction along $x$ and the velocity-gradient direction along $y$.
The initial configurations are prepared by relaxing a purely repulsive system.
Shear is started from a mechanically relaxed configuration for which the total kinetic energy
$\sum_i m_i v_i^2/2N$
is smaller than $1.0\times10^{-8} k_{\rm n}d_0^2N$.
We examine shear rates
$5\times10^{-6}\le\dot\gamma\le5\times10^{-3}$
and packing fraction
$0.730\le\varphi\le0.810$.

The equations of motion are numerically integrated using LAMMPS
\cite{plimpton1995fast,thompson2022lammps}
with a time step
$\Delta t=2.5\times10^{-3}\sqrt{m_0/k_\mathrm{n}}$.
We use $d_0$, $m_0$, and $k_\mathrm{n}$ as the units of length, mass, and spring constant, respectively.
The other parameters are
$k_\mathrm{t}=k_\mathrm{n}$,
$\mu_p=1$,
$\eta_\mathrm{n}=m_{ij}^{\rm red}\gamma_\mathrm{n}$,
and
$\gamma_\mathrm{n}=1.0\sqrt{k_\mathrm{n}/m_0}$,
where
$m_{ij}^{\rm red}=m_i m_j/(m_i+m_j)$
is the reduced mass.
All runs are performed up to an accumulated strain $\gamma=10$, and the quantities reported below are averaged over the common sampling window
$8\le\gamma\le10$.

\begin{figure}[ht]
  \centering
  \includegraphics[width=1.\linewidth]{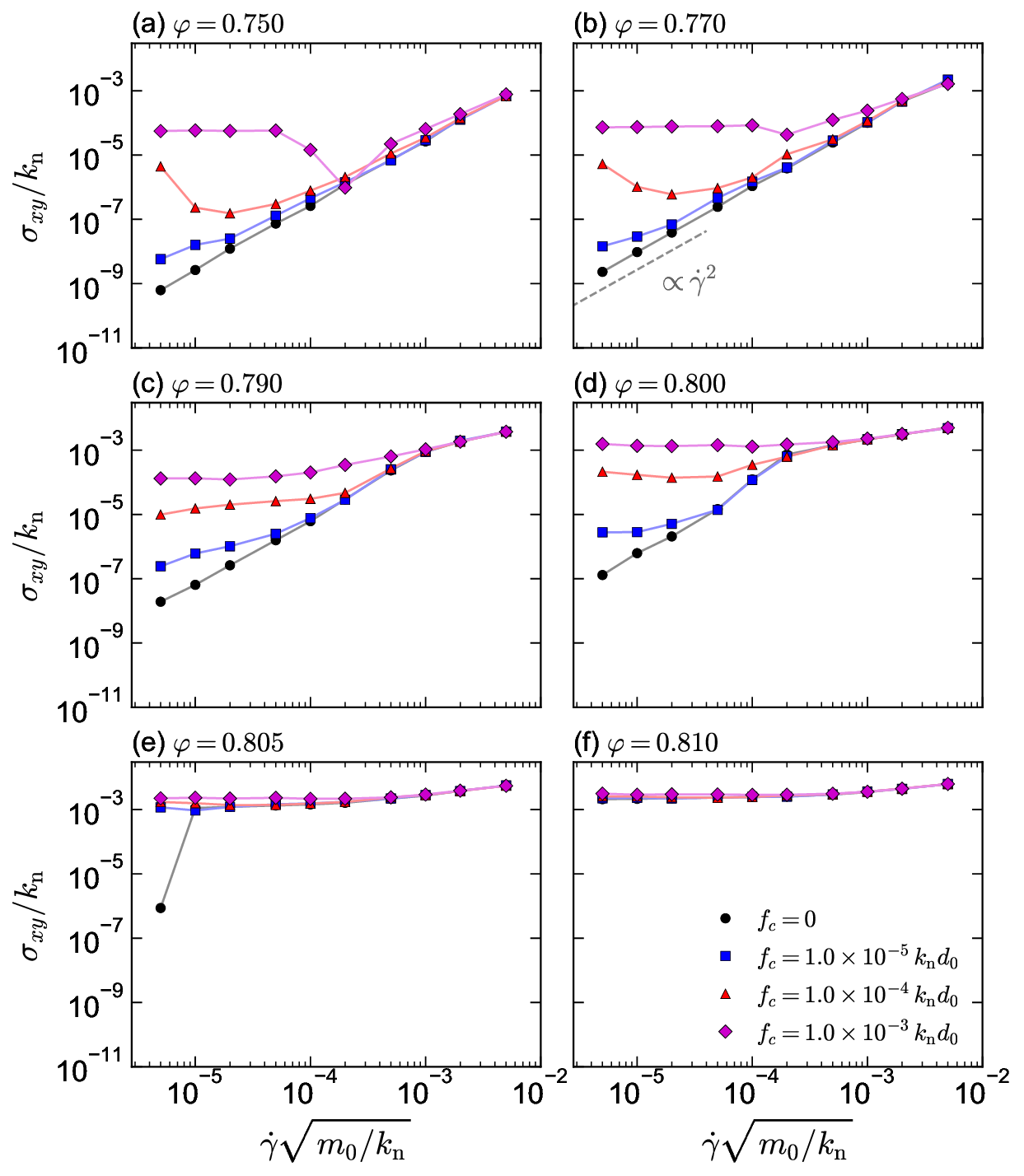}
\caption{
Flow curves $\sigma_{xy}(\dot\gamma)$ for
(a) $\varphi=0.750$,
(b) $0.770$,
(c) $0.790$,
(d) $0.800$,
(e) $0.805$, and
(f) $0.810$.
The symbols represent the attraction strength:
$f_c=0$ (black circles),
$f_c=10^{-5}k_{\rm n}d_0$ (blue squares),
$f_c=10^{-4}k_{\rm n}d_0$ (red triangles), and
$f_c=10^{-3}k_{\rm n}d_0$ (magenta diamonds).
The dashed line in (b) represents the Bagnold scaling
$\sigma_{xy}\propto\dot\gamma^2$.
}
  \label{fig:flow_adh}
\end{figure}

\section{Results}
\label{sec:results}
\subsection{Flow curves and classification of rheological responses}
\label{subsec:adhesive}

Figure~\ref{fig:flow_adh} shows the shear stress as a function of the shear rate.
The shear stress is defined as the sum of the kinetic and virial contributions:
\begin{align}
   \sigma_{\alpha\beta}
   =-\frac{1}{L^2}
   \left[
   \sum_i m_i\delta v_{i,\alpha}\delta v_{i,\beta}
   +\sum_i\sum_{j>i}r_{ij,\alpha}F_{ij,\beta}
   \right],
   \label{eq:stress}
\end{align}
where $\alpha,\beta\in\{x,y\}$ and
$\delta\bm v_i=\bm v_i-\bm V(y_i)$ is the velocity relative to the local mean flow.
Here, $\bm V(y)$ is the local mean velocity coarse-grained along the velocity-gradient direction $y$, and is given by
$\bm V(y)=V_x(y)\bm e_x$ in our system.

For the dry system, the flow curves at
$\varphi=0.75$, $0.77$, and $0.79$
[Figs.~\ref{fig:flow_adh}(a)--(c)]
obey Bagnold scaling,
$\sigma_{xy}\propto\dot\gamma^2$,
over the entire range of shear rates
\cite{bagnold1954experiments,otsuki2011critical}.
At $\varphi=0.805$, the shear stress increases by approximately three orders of magnitude between
$\dot\gamma=5\times10^{-6}$ and $10^{-5}$, indicating DST
[Fig.~\ref{fig:flow_adh}(e)].
In contrast, at $\varphi=0.81$, the system remains on the high-stress branch over the entire range of shear rates, and the stress depends only weakly on $\dot\gamma$
[Fig.~\ref{fig:flow_adh}(f)].
Hereafter, we refer to the low- and high-stress branches as the flowing and thickened branches, respectively.

The effect of attraction appears mainly at low shear rates, while the thickened branch is almost unaffected.
For $\varphi\geq0.80$, the curves for all attraction strengths nearly coincide for
$\dot\gamma\gtrsim10^{-3}$, and at $\varphi=0.81$ they overlap over the entire range of shear rates.
For the weakest attraction, $f_c=10^{-5}$, the stress is slightly enhanced at low shear rates, but the overall shape of the flow curve remains close to that of the dry system.
An exception occurs at $\varphi=0.805$: while the dry system is on the flowing branch at the lowest shear rate, the system with $f_c=10^{-5}$ remains on the thickened branch over the entire range examined.
Thus, even the weakest attraction examined selects the high-stress branch, and the discontinuous jump observed in the dry system is no longer seen under the present rate-controlled protocol.

For $f_c=10^{-3}$, the flowing branch is replaced at low shear rates by a stress plateau, whose magnitude increases with packing fraction.
At $\varphi=0.80$, where the dry system exhibits a sharp increase in stress, the plateau reaches approximately the stress level of the thickened branch and fills the gap between the two branches
[Fig.~\ref{fig:flow_adh}(d)].
Thus, the discontinuous stress increase observed in the dry system is suppressed in two different ways depending on the packing fraction: selection of the high-stress branch and formation of a stress plateau.

At lower packing fractions, the flow curves exhibit an additional nonmonotonic decrease at intermediate shear rates.
For $\varphi=0.75$, the stress decreases by nearly two orders of magnitude from the low-rate plateau at
$\dot\gamma\simeq10^{-4}$--$2\times10^{-4}$,
and then increases again toward the dry flow curve at high shear rates
[Fig.~\ref{fig:flow_adh}(a)].
We refer to this low-stress region as the stress-collapse regime.
The collapse is most pronounced at low packing fractions: only a shallow minimum remains at $\varphi=0.77$, and the nonmonotonic behavior disappears at higher packing fractions.
A similar behavior is observed for $f_c=10^{-4}$, where the stress first decreases and then increases with increasing shear rate.
In this case, however, the minimum occurs at a shear rate approximately one decade lower than that for $f_c=10^{-3}$, and the low-rate side forms a broad shoulder rather than a well-defined stress plateau.

\begin{figure}[t]
  \centering
  \includegraphics[width=0.98\linewidth]{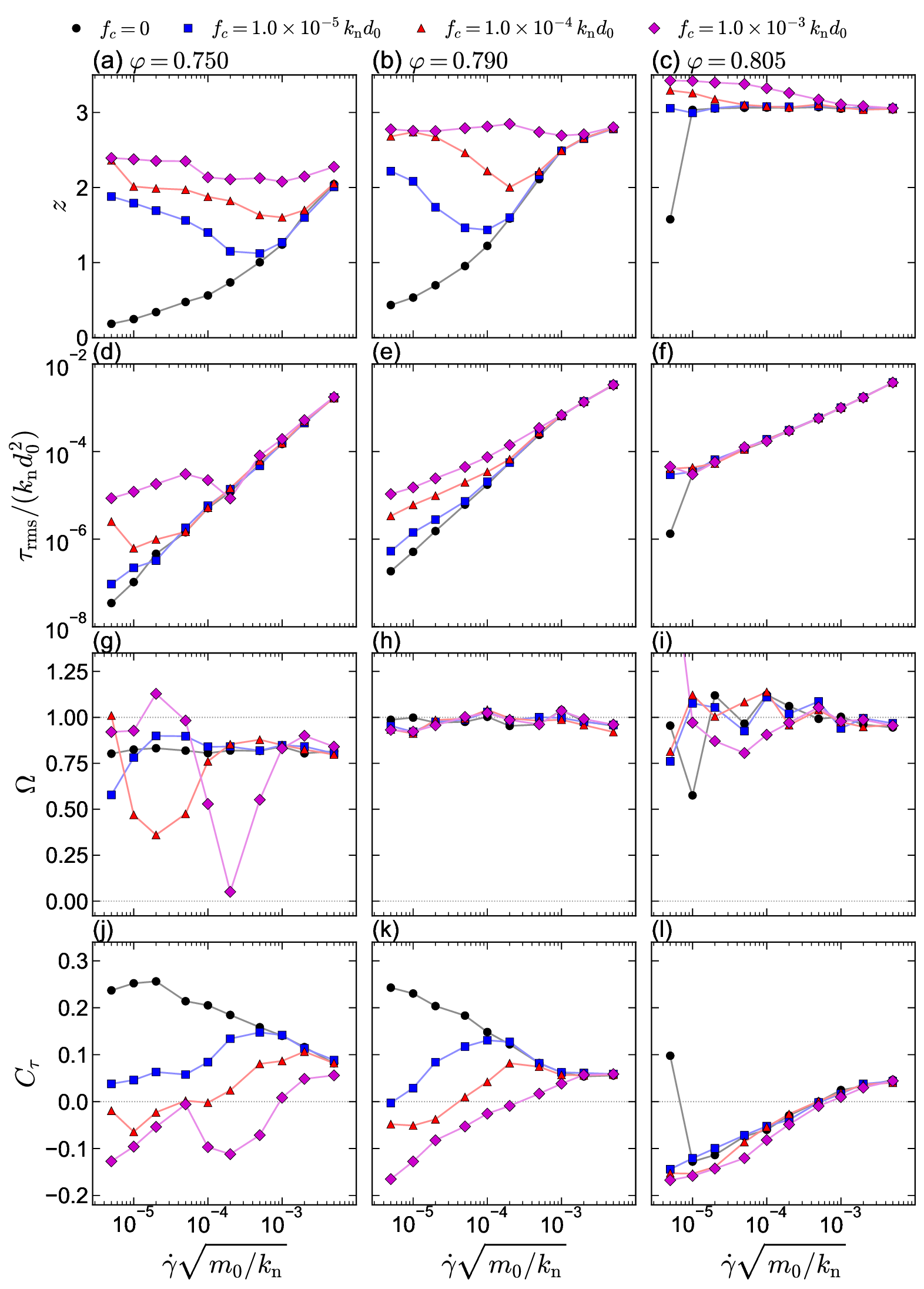}
  \caption{
Shear-rate dependence of the contact and rotational statistics for
$\varphi=0.750$, $0.790$, and $0.805$ (columns).
The symbols are the same as those in Fig.~\ref{fig:flow_adh}.
(a)--(c) Coordination number $z$,
(d)--(f) torque amplitude $\tau_{\rm rms}$,
(g)--(i) normalized mean spin $\Omega$, and
(j)--(l) torque-sign correlation between contacting particles $C_\tau$.
The dotted lines indicate $\Omega=1$ (affine rotation),
$\Omega=0$, and $C_\tau=0$, respectively.
}
  \label{fig:rateresp}
\end{figure}

\subsection{Contact network and rotational states: plateau, branch selection, and collapse}
\label{subsec:rateresp}

Figure~\ref{fig:rateresp} shows the shear-rate dependence of the coordination number $z$, torque amplitude $\tau_{\rm rms}$, normalized mean spin $\Omega$, and torque-sign correlation $C_\tau$ between contacting particles.
The coordination number $z$ is defined as the average number of overlapping contacts, $r_{ij}<d_{ij}$, per particle, and the torque amplitude is measured by
\begin{align}
   \tau_{\rm rms}
   \equiv
   \left\langle \tau_i^2 \right\rangle^{1/2}.
\end{align}
The average particle rotation relative to the imposed shear is characterized by
\begin{align}
   \Omega
   \equiv
   \dfrac{\langle \omega_i \rangle}{-\dot\gamma/2},
   \label{eq:omega}
\end{align}
where $\Omega=1$ corresponds to the affine rotation rate $-\dot\gamma/2$, while $\Omega=0$ indicates a vanishing mean spin.
The local organization of torque signs is characterized by
\begin{align}
   C_\tau
   \equiv
   \big\langle
   \mathrm{sgn}(\tau_i)\,
   \mathrm{sgn}(\tau_j)
   \big\rangle_{\rm contacts},
   \label{eq:ctau}
\end{align}
where $C_\tau>0$ and $C_\tau<0$ indicate preferences for like-sign and opposite-sign torques, respectively.
These quantities exhibit distinct signatures of the stress plateau, high-stress branch selection, and stress collapse observed in Fig.~\ref{fig:flow_adh}.

At low and intermediate packing fractions, attraction strongly increases the coordination number at low shear rates, producing a highly coordinated contact network from the sparse dry network
[Figs.~\ref{fig:rateresp}(a) and (b)].
For example, at $\varphi=0.75$ and $\dot\gamma=5\times10^{-6}$,
$z$ increases from $0.19$ in the dry system to approximately $2.4$ for $f_c=10^{-3}$, while $\tau_{\rm rms}$ increases by approximately two orders of magnitude
[Fig.~\ref{fig:rateresp}(d)].
Thus, the low-rate adhesive plateau is accompanied by a highly coordinated frictional contact network and substantially enhanced frictional torque.

Despite this increase in coordination, the mean spin changes only weakly in the plateau regime.
Except in the collapse regime, $\Omega$ remains close to unity
[Figs.~\ref{fig:rateresp}(g)--(i)],
showing that increased coordination does not suppress mean particle rotation.
The change is more clearly reflected in $C_\tau$
[Figs.~\ref{fig:rateresp}(j)--(l)].
The dry system exhibits $C_\tau>0$ at low shear rates, whereas increasing attraction weakens this positive correlation and produces $C_\tau<0$ for $f_c=10^{-3}$.
Attraction therefore modifies not only the torque amplitude but also the local torque-sign organization.

Near jamming, this reorganization is associated with the high-stress branch selection shown in Fig.~\ref{fig:flow_adh}(e).
For the dry system at $\varphi=0.805$ and $\dot\gamma=5\times10^{-6}$, the system remains on the low-coordination flowing branch with $z\simeq1.6$.
Even the weakest attraction, $f_c=10^{-5}$, increases $z$ to approximately $3.1$, and a highly coordinated state with $z\simeq3$ or larger persists over the entire shear-rate range
[Fig.~\ref{fig:rateresp}(c)].
At the same time, $\tau_{\rm rms}$ increases and $C_\tau$ changes from positive to negative
[Figs.~\ref{fig:rateresp}(f) and (l)].
A similar negative torque-sign correlation is observed on the dry thickened branch.
Thus, the attraction-selected state shares high coordination and negative torque-sign correlations with the dry thickened branch.
Because attraction exerts no direct torque, these changes arise through reorganization of the frictional contact network.

In contrast, the low-density stress-collapse regime exhibits qualitatively different behavior.
For $\varphi=0.75$ and $f_c=10^{-3}$, increasing the shear rate from the low-rate plateau leads to simultaneous decreases in $\tau_{\rm rms}$ and $\Omega$
[Figs.~\ref{fig:rateresp}(d) and (g)].
At $\dot\gamma=2\times10^{-4}$, $\Omega$ decreases to approximately $0.05$, while $z$ remains finite.
The stress collapse therefore occurs with strongly suppressed frictional torque and mean spin despite the persistence of a finite contact network.

At high shear rates, the effect of attraction gradually disappears.
Both $z$ and $\tau_{\rm rms}$ approach their dry values, while $\Omega$ and $C_\tau$ also converge toward the dry response.
This is consistent with the convergence of the attractive and dry flow curves at high shear rates in Fig.~\ref{fig:flow_adh}.

Overall, the three rheological responses are associated with distinct contact and rotational states.
The adhesive plateau is characterized by enhanced coordination and torque, the attraction-selected high-stress state near jamming by high coordination and negative torque-sign correlations, and the stress-collapse state by strongly suppressed torque and mean spin despite a finite contact network.

\begin{figure}[t]
  \centering
  \includegraphics[width=0.95\linewidth]{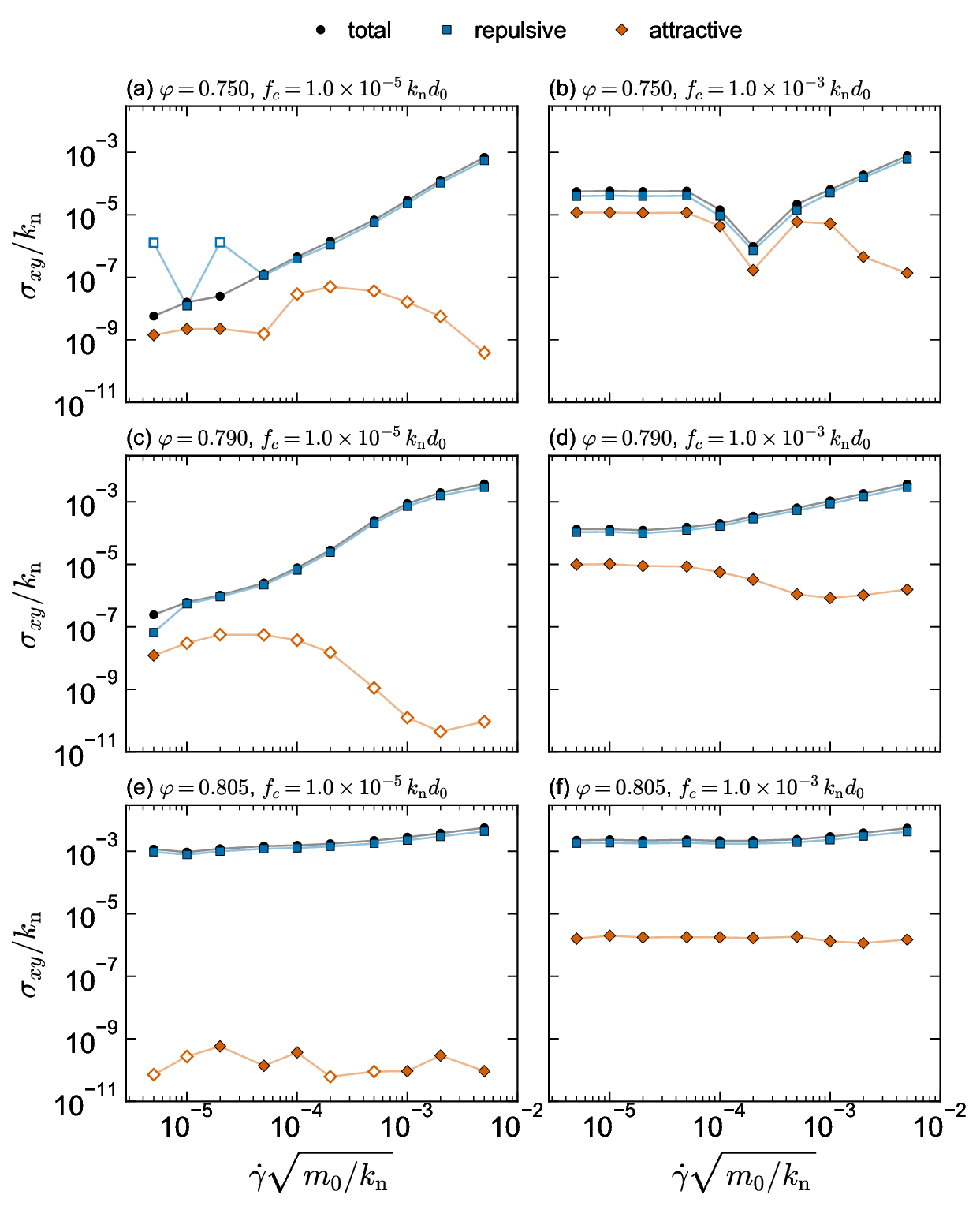}
\caption{
Shear-stress decomposition for
$\varphi=0.750$, $0.790$, and $0.805$ (rows) and
$f_c=10^{-5}k_{\rm n}d_0$ and $10^{-3}k_{\rm n}d_0$ (columns).
Black circles, blue squares, and orange diamonds represent
the total shear stress $\sigma_{xy}$,
the repulsive contribution $\sigma_{xy}^{\rm rep}$,
and the attractive contribution $\sigma_{xy}^{\rm att}$, respectively.
The total stress also includes tangential, normal dissipative, and kinetic contributions.
Open symbols indicate negative values plotted by their absolute magnitudes.
}
  \label{fig:budget}
\end{figure}

\subsection{Shear-stress decomposition}
\label{subsec:budget}

To distinguish the direct stress carried by attraction from changes in stress transmission through the contact network, we decompose the virial contribution of the conservative normal force into the repulsive and attractive branches,
$\sigma_{xy}^{\rm rep}$ and $\sigma_{xy}^{\rm att}$, respectively.
Figure~\ref{fig:budget} compares these contributions with the total shear stress $\sigma_{xy}$.
The total stress also contains contributions from the tangential force, normal dissipation, and kinetic term; for representative dense conditions, the latter two are each less than $1\%$ of the total stress.

At low and intermediate packing fractions
[Figs.~\ref{fig:budget}(a)--(d)],
the attractive branch carries a finite shear stress at low rates, but accounts for only part of the total stress.
For example, at $\varphi=0.75$, $\dot\gamma=5\times10^{-6}$, and
$f_c=10^{-3}$, the total stress is
$5.6\times10^{-5}$, whereas
$\sigma_{xy}^{\rm att}\simeq1.2\times10^{-5}$
[Fig.~\ref{fig:budget}(b)].
Instead, the repulsive contribution is strongly enhanced as the adhesive plateau develops.
The relative attractive contribution becomes even smaller at $\varphi=0.79$
[Figs.~\ref{fig:budget}(c) and (d)].
Thus, the adhesive plateau cannot be understood as a simple addition of attractive virial stress to the dry-system stress; attraction primarily modifies stress transmission through the contact network, consistent with the increases in $z$ and $\tau_{\rm rms}$ shown in Fig.~\ref{fig:rateresp}.

Near jamming, this distinction is particularly clear
[Figs.~\ref{fig:budget}(e) and (f)].
The direct attractive contribution remains very small, while the total stress is dominated by the repulsive contribution.
Nevertheless, even the weakest attraction selects the high-stress branch under conditions where the dry system remains on the low-stress branch
[Fig.~\ref{fig:flow_adh}(e)].
Thus, high-stress branch selection is associated not with attraction directly carrying the stress, but with attraction selecting a highly coordinated frictional contact state with repulsion-dominated stress transmission.

The stress-collapse regime provides the complementary case.
At $\varphi=0.75$ and $f_c=10^{-3}$, both the total and repulsive stresses decrease strongly at intermediate shear rates, whereas the attractive contribution remains finite
[Fig.~\ref{fig:budget}(b)].
Together with the finite coordination number in Fig.~\ref{fig:rateresp}(a), this shows that neither attraction nor the presence of contacts alone is sufficient to sustain a large macroscopic shear stress.

Overall, Figs.~\ref{fig:budget}(a)--(f) show that attraction makes a finite direct contribution to the shear stress, but the major rheological changes are associated with reorganization of the stress-carrying contact network.
At low rates this reorganization enhances repulsion-dominated stress transmission, near jamming it selects the high-stress branch, whereas in the low-density collapse regime stress transmission is strongly suppressed despite a finite contact network.

\begin{figure*}[htb]
  \centering
  \includegraphics[width=0.85\linewidth]{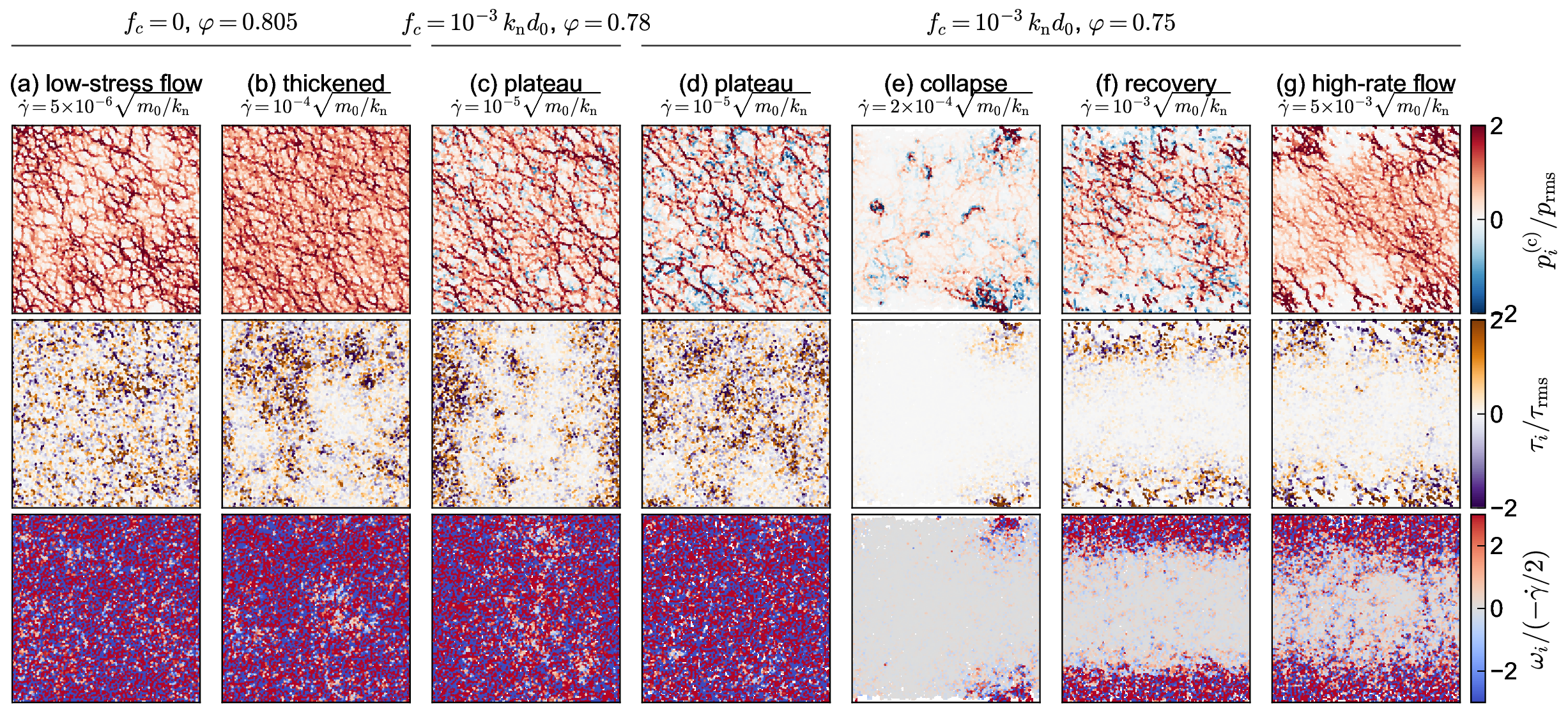}
\caption{
Snapshots of seven representative states.
(a) Low-stress flow at $f_c=0$ and $\varphi=0.805$,
(b) the thickened state at the same $f_c$ and $\varphi$,
(c) the adhesive plateau at $f_c=10^{-3}k_{\rm n}d_0$ and $\varphi=0.78$,
and (d)--(g) the plateau, stress collapse, recovery, and high-rate flow,
respectively, with increasing shear rate at
$f_c=10^{-3}k_{\rm n}d_0$ and $\varphi=0.75$.
The top row shows the normalized particle pressure
$p_i^{\rm (c)}/p_{\rm rms}$ (red: compression; blue: tension),
the middle row the normalized particle torque
$\tau_i/\tau_{\rm rms}$,
and the bottom row the normalized spin
$\omega_i/(-\dot\gamma/2)$
(white indicates weak particle rotation).
}
  \label{fig:torquemap}
\end{figure*}

\subsection{Real-space organization and single-particle statistics}
\label{subsec:states}

We next examine how the representative flow states differ
in the real-space distributions of particle pressure, torque, and spin.

We characterize the local compressive and tensile states
associated with the conservative normal force
using the particle pressure $p_i^{\rm (c)}$.
By assigning half of each pair virial to each particle,
we define the virial stress of particle $i$ as
\begin{align}
   \sigma^{\rm (c)}_{i,\alpha\beta}
   =
   -\frac{2}{\pi d_i^2}
   \sum_{j\neq i}
   r_{ij,\alpha}
   F^{\rm (c)}_{ij,\beta},
\end{align}
and its isotropic component as
\begin{align}
   p_i^{\rm (c)}
   \equiv
   -\frac{1}{2}
   {\rm Tr}\,\bm{\sigma}_i^{\rm (c)}.
\end{align}
Thus, $p_i^{\rm (c)}>0$ represents net compression,
whereas $p_i^{\rm (c)}<0$ represents net tension.
In the snapshots below,
$p_i^{\rm (c)}$ is normalized by
$p_{\rm rms}\equiv\langle (p_i^{\rm (c)})^2\rangle^{1/2}$
for each state.

Figure~\ref{fig:torquemap} compares the dry low-stress flow
and thickened state, the adhesive plateau,
and the collapse and subsequent recovery
in the low-density attractive system.
In the dry low-stress flow, dry thickened state,
and adhesive plateau
[Figs.~\ref{fig:torquemap}(a)--(d)],
force-bearing contact structures extend throughout the system,
and substantial particle rotation is observed.
Their torque organization, however, differs.
Like-sign torques tend to occur locally in the dry low-stress flow,
consistent with $C_\tau>0$,
whereas positive and negative torques are more locally intermixed
in the dry thickened state and on the adhesive plateau,
where $C_\tau<0$.
The attractive states are further distinguished
by the appearance of particles under net tension
[Figs.~\ref{fig:torquemap}(c),(d)].

The stress-collapse state
[Fig.~\ref{fig:torquemap}(e)]
shows a qualitatively different spatial organization.
Compressed and tensile particles coexist,
the torque field becomes strongly heterogeneous,
and appreciable particle spin is confined to a narrow region.
Thus, the suppression of $\tau_{\rm rms}$ and $\Omega$
seen in Fig.~\ref{fig:rateresp}
is accompanied by strong spatial localization.
With increasing shear rate
[Figs.~\ref{fig:torquemap}(f),(g)],
torque and spin again spread over broader regions,
indicating recovery toward the high-rate flow.

\begin{figure*}[t]
  \centering
  \includegraphics[width=0.98\linewidth]{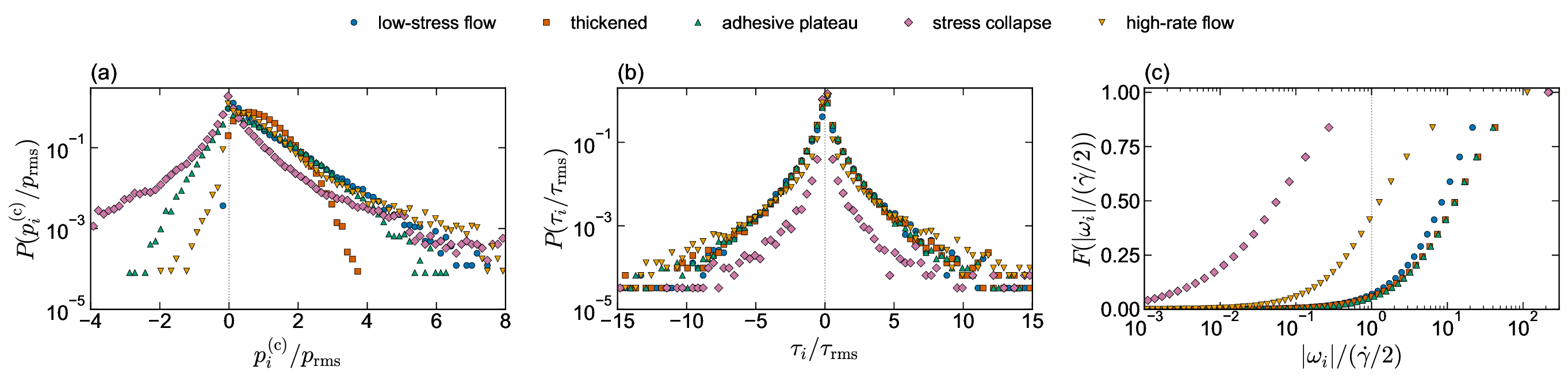}
  \caption{
Single-particle statistics for five representative states.
The low-stress flow and thickened state correspond to
$f_c=0$, $\varphi=0.805$, and
$\dot\gamma\sqrt{m_0/k_{\rm n}}=5\times10^{-6}$ and $10^{-4}$, respectively.
The adhesive plateau corresponds to
$f_c=10^{-3}k_{\rm n}d_0$, $\varphi=0.78$, and
$\dot\gamma\sqrt{m_0/k_{\rm n}}=10^{-5}$;
the stress collapse to
$f_c=10^{-3}k_{\rm n}d_0$, $\varphi=0.75$, and
$\dot\gamma\sqrt{m_0/k_{\rm n}}=2\times10^{-4}$;
and the high-rate flow to
$f_c=10^{-3}k_{\rm n}d_0$, $\varphi=0.75$, and
$\dot\gamma\sqrt{m_0/k_{\rm n}}=5\times10^{-3}$.
(a) Distribution of the normalized particle pressure
$p_i^{\rm (c)}/p_{\rm rms}$, where negative values indicate tension;
(b) distribution of the normalized torque $\tau_i/\tau_{\rm rms}$;
and (c) cumulative distribution of the normalized spin amplitude
$|\omega_i|/(\dot\gamma/2)$.
The dotted line in (c) indicates the affine rotation rate,
$|\omega_i|/(\dot\gamma/2)=1$.
}
  \label{fig:distributions}
\end{figure*}

Figure~\ref{fig:distributions} provides a statistical comparison
of these representative states.
The particle-pressure distributions
[Fig.~\ref{fig:distributions}(a)]
show that attraction extends the distribution toward negative pressure,
consistent with the tensile particles visible in Fig.~\ref{fig:torquemap}.
The collapse exhibits substantial weight
on both the compressive and tensile sides.

When the particle torque is normalized by $\tau_{\rm rms}$,
the distribution shapes differ only weakly among the flowing states
[Fig.~\ref{fig:distributions}(b)].
Thus, the principal changes in torque statistics
are expressed more strongly in the absolute torque scale $\tau_{\rm rms}$
and in the torque-sign correlation $C_\tau$
than in the normalized one-particle distribution.

The distinction between the adhesive plateau and the collapse
is particularly clear in the spin distributions
[Fig.~\ref{fig:distributions}(c)].
The plateau retains a broad distribution of particle spins,
whereas the collapse shifts strongly toward small rotation,
with the median of $|\omega_i|/(\dot\gamma/2)$ decreasing to $0.06$.
The decrease in $\Omega$ therefore reflects
a suppression of the individual spin amplitudes
rather than cancellation between positive and negative spins.

Together, Figs.~\ref{fig:torquemap} and \ref{fig:distributions}
show that the dry low-stress flow, dry thickened state,
and adhesive plateau remain rotationally active
while differing in torque-sign organization and local pressure statistics.
The stress-collapse state is distinct,
with strongly suppressed and spatially localized torque and particle rotation.

\begin{figure*}[t]
  \centering
  \includegraphics[width=0.75\linewidth]{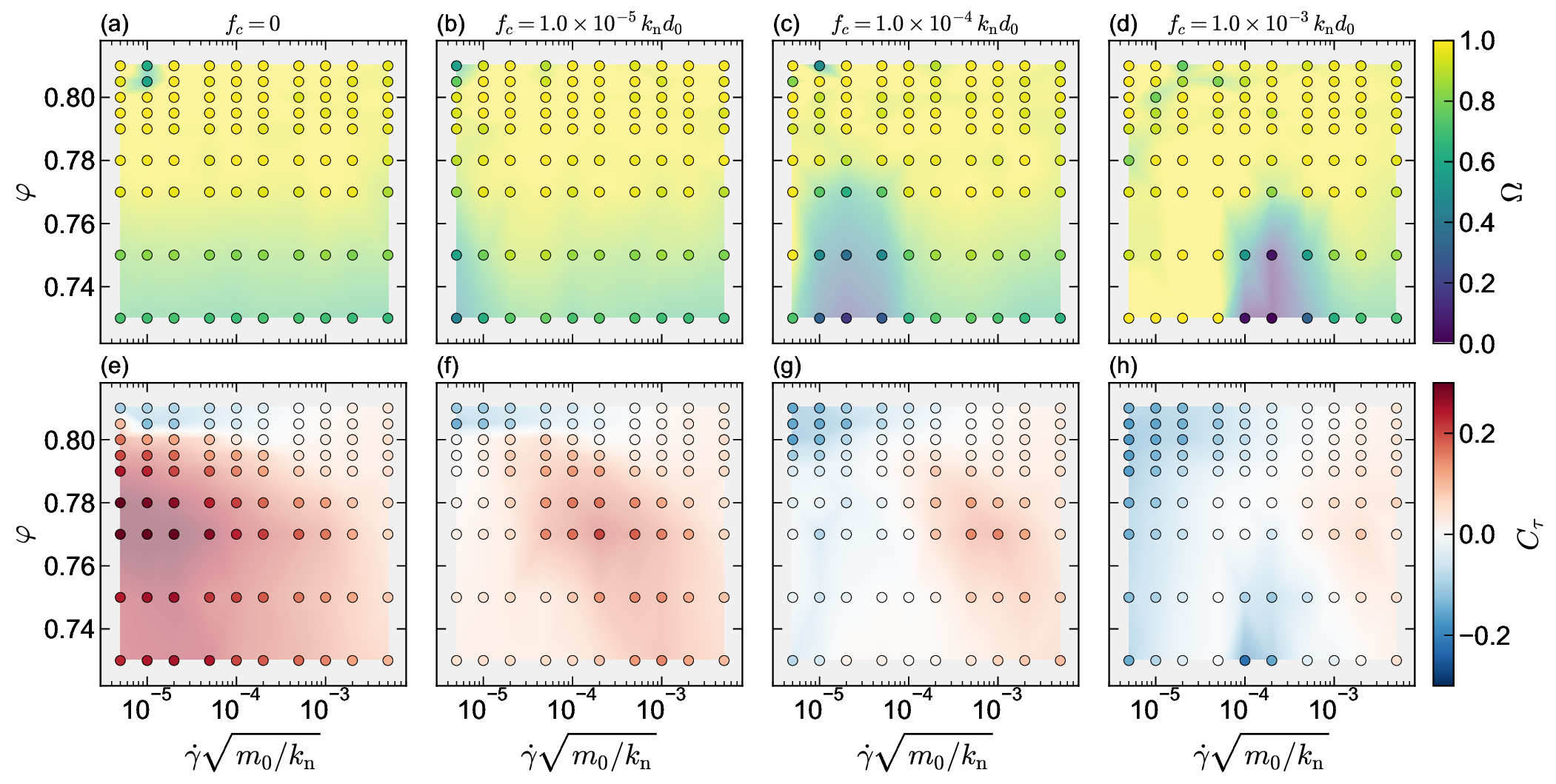}
  \caption{
State maps of rotational and torque statistics
in the $(\dot\gamma,\varphi)$ plane.
Columns correspond to
$f_c=0,\,10^{-5},\,10^{-4},\,10^{-3}$.
(a)--(d) Normalized mean spin $\Omega$;
(e)--(h) torque-sign correlation $C_\tau$
between contacting particles.
}
  \label{fig:phasemap}
\end{figure*}

\subsection{State maps}
\label{subsec:map}

We have so far examined the adhesive plateau,
high-stress branch selection,
and stress-collapse state
at representative packing fractions and shear rates.
We now summarize where the corresponding rotational
and torque states occur in the $(\dot\gamma,\varphi)$ plane.
Figure~\ref{fig:phasemap} shows
the normalized mean spin $\Omega$
and the torque-sign correlation $C_\tau$
for each attraction strength.

The mean spin $\Omega$ remains close to unity
over most of the parameter space
[Figs.~\ref{fig:phasemap}(a)--(d)].
Thus, even when attraction strongly modifies
the shear stress and contact state,
the mean particle rotation remains close
to the affine rotation rate in most flowing states.
For the strongest attraction studied, however,
a region of sharply reduced $\Omega$
appears at low packing fractions
and intermediate shear rates
[Fig.~\ref{fig:phasemap}(d)].
This region corresponds to the stress-collapse state
identified in Fig.~\ref{fig:rateresp}
and visualized in Fig.~\ref{fig:torquemap}(e).
For weaker attraction
[Figs.~\ref{fig:phasemap}(b),(c)],
rotational suppression is also observed at low packing fractions,
but occurs at lower shear rates
and is less pronounced.
Thus, $\Omega$ provides a clear map
of the rotational suppression associated with the collapse.

The behavior of $C_\tau$ captures a different aspect
of the flow states
[Figs.~\ref{fig:phasemap}(e)--(h)].
In the dry system,
positive torque-sign correlations $C_\tau>0$
occur in the low-rate flows
at low and intermediate packing fractions,
whereas $C_\tau<0$ is found
in the high-density thickened state
[Fig.~\ref{fig:phasemap}(e)].
With increasing attraction,
the region of $C_\tau<0$ extends toward lower shear rates,
and for strong attraction
negative local torque-sign correlations occur
over a broad range of packing fractions
[Figs.~\ref{fig:phasemap}(f)--(h)].
This is consistent with
Figs.~\ref{fig:rateresp} and \ref{fig:torquemap},
where the adhesive plateau
shares negative local torque-sign correlations
with the dry thickened state.

Importantly, $C_\tau<0$ is not specific
to the stress-collapse state.
Negative correlations also occur
in the dry thickened state and on the adhesive plateau,
where $\Omega$ remains close to unity.
In the collapse regime, by contrast,
negative $C_\tau$ is accompanied
by a pronounced decrease in $\Omega$.
As shown by the single-particle spin distributions
in Fig.~\ref{fig:distributions}(c),
this reduction in $\Omega$
reflects a suppression of individual particle-spin amplitudes
rather than cancellation between positive and negative spins.

Thus, $C_\tau$ and $\Omega$
characterize different aspects of the attraction-induced state changes.
The torque-sign correlation $C_\tau$ probes
the local organization of torque signs
between contacting particles within flowing states,
whereas $\Omega$ captures the suppression
of mean particle rotation associated with the stress collapse.
Together, these rotational and torque statistics
provide a microscopic classification of the flow states
that is complementary to the macroscopic shear stress.

\section{Discussion and summary}
\label{sec:discussion}

The central result of this study is that a short-range central attraction,
which exerts no direct torque on the particles,
strongly modifies torque organization, particle rotation,
and shear-stress transmission through reorganization of the frictional contact network.
At low shear rates, attraction increases the coordination number and torque amplitude,
while the torque-sign correlation $C_\tau$ changes from positive in the dry low-rate flow
to negative in the adhesive plateau.
The shear-stress decomposition supports the same picture:
although the attractive branch carries a finite shear stress,
its direct contribution accounts for only part of the total stress,
whereas the repulsive contribution is strongly enhanced.
Near jamming, even weak attraction selects the high-stress branch
despite its small direct stress contribution.
Thus, the dominant effect of attraction is not simply to add attractive virial stress,
but to modify the frictional contact network through which stress and torque are transmitted.

This network reorganization produces different macroscopic responses
depending on packing fraction and shear rate.
At low shear rates, attraction produces an adhesive stress plateau,
while near the dry-system DST regime it selects the high-stress branch.
At low packing fractions and intermediate shear rates,
a qualitatively different stress-collapse state appears.
There, the coordination number remains finite,
but the shear stress, torque amplitude, and particle rotation are strongly suppressed,
together with pronounced spatial heterogeneity and coexistence of particles
under net compression and tension.
The collapse therefore cannot be interpreted as a simple loss of contacts;
rather, it corresponds to a contact state that transmits shear stress
and frictional torque inefficiently.
Whether the high-stress branch selection itself represents a discontinuous transition,
and where its threshold lies,
cannot be determined from the present spacing of attraction strengths.

The rotational statistics provide information that is not contained
in the macroscopic shear stress alone.
The dry thickened state and the adhesive plateau differ in stress scale
and in the presence of tensile particles,
but share high coordination, substantial substantial spin fluctuations,
and negative local torque-sign correlations.
By contrast, the dry low-rate flow exhibits $C_\tau>0$.
This is consistent with the disappearance of like-torque clusters across dry DST
reported in ref.~\cite{rahbari2021fluctuations},
while the present results show that attraction extends the region of negative
torque-sign correlations toward lower shear rates.
The stress-collapse state is distinguished instead by a pronounced reduction in $\Omega$.
The single-particle spin distributions show that this decrease reflects
suppression of the spin amplitude itself rather than cancellation of opposite spins.
Thus, $C_\tau$ characterizes the local organization of frictional torques,
whereas $\Omega$ provides a complementary measure of rotational suppression.

In summary, short-range attraction can control rotational and rheological states
without directly exerting particle torque.
By reorganizing the frictional contact network,
it produces an adhesive stress plateau,
selects a high-stress state near jamming,
and generates a distinct low-density stress-collapse regime.
The stress decomposition and rotational statistics together show that
these changes cannot be understood from the direct attractive stress
or the macroscopic flow curve alone.
Further work should establish the robustness and boundaries of these states
and clarify their microscopic origin through contact anisotropy,
friction mobilization, and contact-network topology~\cite{d2025topological}.

\section*{ACKNOWLEDGMENTS}
The numerical simulations were mainly conducted using the JAXA Supercomputer System Generation 3 (JSS3).
K.Y. is partially supported by JSPS KAKENHI (Grant Nos.~24KJ0110, 25K01063, and 26K06964).



\section*{DATA AVAILABILITY}
The data that support the findings of this study will be deposited in Zenodo upon acceptance.

\renewcommand{\bibliography}[1]{}

\end{document}